\documentclass[preprint,12pt]{elsarticle}

\usepackage{graphicx} 
\usepackage{amssymb}

 \usepackage{lineno}

\usepackage{color}

\usepackage{hyperref}
\usepackage[numbers]{natbib}
\journal{Book: Smart Cities, Cybersecurity and Privacy}

\begin{document}

\begin{frontmatter}


\title{Crowdsensing and privacy in smart city applications}



\author[csiro]{Raj Gaire}
\author[iit-k]{Ratan K Ghosh}
\author[csiro]{Jongkil Kim}
\author[csiro]{Alexander Krumpholz}
\author[uni-nc]{Rajiv Ranjan}
\author[iit-b]{RK Shyamasundar}
\author[csiro]{Surya Nepal}

\address[csiro]{CSIRO, Australia}
\address[uni-nc]{University of Newcastle, UK}
\address[iit-k]{IIT Kanpur, India}
\address[iit-b]{IIT Bombay, India}

\begin{abstract}
Smartness in smart cities is achieved by sensing phenomena of interest and using them to make smart decisions.
Since the decision makers may not own all the necessary sensing infrastructures, crowdsourced sensing, can help collect important information of the city in near real-time.
However, involving people brings of the risk of exposing their private information.This chapter explores crowdsensing in smart city applications and its privacy implications.
\end{abstract}


\end{frontmatter}


\section{Introduction}
\label{S:introduction}
Smart cities are intelligent cities.
Smartness of economy, people, governance, mobility, environment and living are the defining characteristics of smart cities~\cite{Giffinger2007}.
Intelligence in a smart city is built upon measuring phenomena or \emph{things} of interest and using them for making smart decisions.
Measuring things using sensors has been considered as the key aspect of a smart city~\cite{caragliu2011smart}.
Today, sensors can measure a wide variety of phenomena.
Moreover, since the size of these sensors has become smaller, they have now been embedded in many household and personal devices including but not limited to smartphones, vehicles, televisions and gaming devices.
These sensors can be used collectively for community based \emph{crowdsourcing} of the collection of measurements, a.k.a.\emph{mobile crowdsensing}~\cite{Ganti2011}.
Since crowdsensing involves people and their private devices, privacy and security are the \emph{prima facia} concerns.
Although the term mobile crowdsensing and crowdsensing are often used interchangeably, in our opinion the term \emph{mobile crowdsensing} covers only a subset of the more broader term \emph{crowdsensing}.
In this chapter, firstly, we will attempt to define crowdsensing as a broader term for crowdsourced sensing.
Secondly, we will discuss privacy issues, concerns and considerations in crowdsensing applications.
Thirdly, we will present two case studies to illustrate privacy issues in smart city crowdsensing and approaches to address those issues.
Finally, we will discuss and draw conclusions.

\section{Defining crowdsensing}
\label{S:crowdsensing}

In this section we will discuss about different terminologies used in sensing domain leading to a generalised definition of crowdsensing.

\subsection{Sensing}
\label{SS:sensing}

The Webster online dictionary defines sensing as \emph{becoming aware of something via the senses}\footnote{http://www.webster-dictionary.org/definition/sensing}.
This is a generic definition.
Note that this definition does not differentiate among different types of senses.
Indeed, sensing can be performed by not only the sensory organs of humans or animals but also different types of sensors including electronic sensors.

In the computer science community, the W3C SSN ontology~\cite{lefort2011semantic} defines sensing as a process that provides an estimated value of a phenomenon.
This definition shows the realisation that measuring a phenomena often involves measurement errors.
Therefore a measurement value should be treated as an estimation rather than an absolute value.
Again, this definition does not differentiate whether the sensing is performed by an electronic device or a non-electronic object including human.

\subsection{Sensors and Sensing}
\label{SS:sensors-sensing}

The W3C SSN ontology~\cite{lefort2011semantic} also defines sensors as things that perform sensing by transforming an incoming stimulus into digital representation.
Low cost of electronic sensors has made them the technology of choice for estimating the value of a phenomenon.
These sensors are already used in modern cities for various purposes.
For example, sensors are used to measure weather conditions such as temperature, relative humidity and wind direction/velocity.
These measurements are used to forecast the fire danger ratings of bushfires in Australia~\cite{aisrf:noble1980},
as well as,
to assess and plan for the fire-fighting activities during bushfires as discussed later in Section \ref{S:casestudy1}.

Smartphones have become ubiquitous in our everyday lives.
The sensors embedded in smartphones can also observe several phenomena.
In addition to the sensors in smartphones, people are now at the forefront of sensing and transforming the observations into electronic data.
In other words, people are working as social sensors~\citep{Roitman:2012kw}.
These observations gathered from social sensors can be used to generate different intelligence.
For instance, people often use social media platform like twitter to publish disaster related observations.
Such data has already been used for early detection of disasters~\cite{aisrf:power}, as well as, 
for the situation awareness during disaster events~\cite{aisrf:yin2012}.

Arguably, human can sense many phenomena for which sensors might be either unavailable or available at a very high cost.
For example, people can assess non-physical phenomena such as feeling, mood, taste and smell.
Furthermore, for some phenomena, the same sensor reading might have different meanings for different people.
For example, the same temperature could be considered as a comfortable temperature to some people while not so comfortable to others.
Therefore, it is important to consider both electronic as well as non-electronic sensing when developing application for smart cities.

\subsection{Crowdsensing in Smart Cities}
\label{SS:crowdsensing}
In case of a smart city, sensing is not limited to a single phenomena.
Rather, it has to cover different phenomena at different locations across the city.
Owning the infrastructure to collect measurements about these phenomena may at times be very expensive and at other times almost impossible.
Therefore, community based crowdsourced sensing, or crowdsensing in short, can play an important role in collecting information in smart cities.

We noticed that the terms \emph{crowdsensing} and \emph{mobile crowdsensing} have often been used interchangeably and we disagree with such uses for the reasons described below.
The specialized term mobile crowdsensing was coined by \citet{Ganti2011} to refer to the crowd based sensing using electronic sensors embedded in personal mobile devices.
\citet{guo2015mobile} extended this concept of mobile crowdsensing to \emph{Mobile Crowd Sensing and Computing (MCSC)}.
Formally, MCSC is defined as 
\emph{``a new sensing paradigm that empowers ordinary citizens to contribute data sensed or generated from their mobile devices 
and aggregates and fuses the data in the cloud for crowd intelligence extraction and human-centric service delivery"}.
This definition includes data sensed by sensors as well as that contributed by users through mobile devices.
Even though the definition of MCSC tries to broaden the notion of mobile crowdsensing, using this definition as a definition of a broader term crowdsensing is still not appropriate for the following reasons.

First, mobility is a major component of the mobile crowdsensing and the MCSC definitions.
However, in many cases, personal sensing devices might not be mobile at all.
For example, homes are often installed with security camera devices at fixed locations that have already been used to sense criminal activities in modern cities.
Similarly, many homes in modern cities have been installed with solar panels for electricity generation that are capable of monitoring the generation and use of electricity in these homes.
These devices are not mobile, yet this information about electricity generation and use combined with weather forecast could be used to predict the electricity demand of a smart city.
We have previously studied the use of locally installed weather stations in farms for welfare assessment of the animals in the farms~\cite{taylor2013farming}.
Similar to farms, people in smart cities could install their own mini weather stations at their backyards to collect and share more precise weather condition data around their homes.
In all the above scenarios, sensors are not attached to any mobile devices.
They are not mobile either.
Yet, they contribute to the intelligence of a smart city.

And second, electronic sensors are the primary components of the above definition of mobile crowdsensing.
However, many of the phenomena of interest need subjective assessment by human.
For example, people constantly sense their environment (e.g.confort, happiness) and share the information in social media such as facebook and twitter using mobile and non-mobile devices.
In a city, people might identify peculiar smell around certain location, or peculiar taste of water at their homes. The would notify to authorities and might also post messages in social media about these issues.
Crowd based sensing mechanisms are already embedded in modern cities as a key component of the post market surveillance of medicine for discovery of their side-effects~\cite{karimi:2015}.
In all these examples, the information can be useful to authorities to identify problems, develop an appropriate solutions quickly and potentially save lives.
These non-mobile data sensed and shared by the crowd, not necessarily using mobile phones, would not be considered as crowdsensed data if judged by the narrow definition of mobile sensing.
Therefore,
a more general definition is required.


\subsection{Defining crowdsensing}
We define crowdsensing as \emph{an inverse form of crowdsourcing~\cite{Brabham:2008jh} in which a vast number of independently owned entities (a.k.a. crowd including but not limited to people) knowingly (e.g.participatory) or unknowingly (e.g.opportunistically) as well as directly (e.g.by typing) or indirectly (e.g.through computer/mobile applications) sense and share the estimation of a phenomena that can be opportunistically used for decision making}.
Unlike in crowdsourcing, the crowd is not asked to contribute data to solve a given problem.
Rather, the crowd senses and shares data which is used opportunistically to make evidence based, and hence smart, decisions.
Internet of things (IoT), specialized applications including prediction applications, mobile phone apps and Internet based applications as well as generic social media applications can be considered as the enablers of crowdsensing.
Considering this broader definition and enablers of crowdsensing is more useful when developing a smart city application as discussed in later sections.

\section{Privacy in Crowdsensing}
\label{S:privacy}

Privacy is a fundamental human need.
As highlighted by \citet{OHara:2011}, a person may feel comfortable to know that 
some information about it are known to friends, others to its banker and yet others to its doctor.
However, the person may not feel comfortable if either of the three know all the facts about it.
This is because the collection of information about an individual allows extraction of knowledge about the individual's habits, beliefs and health and therefore may create unfair disadvantage to the individual~\cite{Cohen:2012, Solove:2011, Magi:2011}.
Furthermore, knowledge of personal facts may lead to inflicting personal harm as well as illegal activities such as identity theft, blackmailing and burglaries against the person.
Therefore, considering privacy, specifically in crowdsensing applications, is very important~\cite{Cilliers:2014}.

According to \citet{MartinezBalleste:2013}, there are five types of citizen's privacy:
Identity Privacy,
Query Privacy,
Location Privacy,
Footprint Privacy, and,
Owner Privacy.
The \emph{identity privacy} describes the problem that users can be identified when they communicate with smart city components.
For example, 
an application installed in a smart phone could not only contribute to crowdsensing but also share the personal information.
The identity privacy is not limited to the identify of the individual contributing to crowdsensing.
Sometimes, the information contributed by a person may contain the private details of other people.
For example, a non-social media user can be profiled using information shared about the person from other users~\cite{Garcia:2017ch}.
Similarly, multimedia data may contain images of other people revealing their privacy~\cite{LiY:2017}.
The \emph{query privacy} relates to the queries asked by users.
By analysing the query, the user could be identified~\cite{jones:2007know}.
The \emph{location privacy} covers the information about the position of users at give times.
Devices like modern mobile phones can gather GPS-based position information like longitude and latitude
revealing spatio-temporal preferences of their users.
Someone querying for a location-based service, e.g. a nearby restaurant, can exposes its location.
From location data and timestamps, a person's demographic information, home and work addresses, commute routes and other habits can be derived~\cite{LiH:2016, Joglekar:2017wb,To:2017vw} 
The \emph{footprint privacy} is about the risks involved with the combination of little pieces of information that are left in a system.
For example, when using a web browser to access a web page, the cookie left on a device is a footprint of the web page that can reveal individual's preferences.
Finally, the \emph{owner privacy} addresses the problem of querying data about the owner of the data contributed by a user.
For example, the citizen may contribute to the electricity use of their home which could be used to infer potentially business sensitive information about the electricity providers. 

The Aadhaar case study presented in Section~\ref{S:casestudy-aadhaar} will illustrate how a system that can potentially bring smartness in an entire country may inflict some of the above mentioned privacy issues.
Legal instruments are necessary to protect citizen privacy as well as to deter any misuse of people's private data.

\subsection{Privacy Laws in Australia}
\label{SS:privacy-law-australia}

In Australia, the Privacy Act 1988\footnote{https://www.legislation.gov.au/Series/C2004A03712} regulates handling of personal information about individuals.
It defines personal information as the \emph{``…information or an opinion, whether true or not, and whether recorded in a material form or not, about an identified individual, or an individual who is reasonably identifiable''}.
Accordingly to this definition, privacy is not only related to individual’s personal information such as name, date of birth, address etc.but also related to any commentary or opinion about the person.
The Privacy Act includes 13 Australian Privacy Principles (APP)\footnote{https://www.oaic.gov.au/individuals/privacy-fact-sheets/general/privacy-fact-sheet-17-australian-privacy-principles} which are applicable to APP entities including most Australian government agencies and businesses.
In addition, the Australian Information Commissioner can make or approve legally binding and non-binding guidelines and rules.

Modern technologies make the collection and storage of data, the extraction of information and the discovery of knowledge fast.
On one hand, it makes new applications like smart cities feasible.
But on the other hand, it can lead to erosion of privacy of the individuals~\cite{Enserink:2015bz}.
Specifically in case of crowdsensing applications, the private information about an individual could be collected or inferred leading to personal, social and legal consequences.
Therefore, smart cities must consider the privacy related obligations and implications of their applications.

\subsection{Privacy obligations}
\label{SS:privacy-obligations}

Using crowdsourced data carries risk of illegally accessing, storing, sharing and potentially revealing private or confidential information about people.
If not considered carefully, the information that should be protected may get released to public not only causing damage to reputation but also leading to legal infringements.
Following are some privacy related obligations for an entity collecting crowdsensed data.

\paragraph{Security of personal information} An entity holding personal information is required to take all reasonable steps to protect the information from misuse, interference and loss, as well as from unauthorised access, modification and disclosure.
The technologies related to privacy and security improve over time.
As such, the steps taken in the past as reasonable measures may not be reasonable in today's context.
Therefore, APP entities should regularly assess their approaches in regards to the security of personal information.

\paragraph{Compliance implications} An APP entity is required to take reasonable steps in terms of implementation practices, procedures and systems to ensure the compliance with APP.
For the reasons similar to above, the APP entities should regularly assess their privacy compliance practices and procedures.

\paragraph{Privacy policy} The APP privacy policy states that an APP entity must clearly express and keep its privacy policy about the management of personal information up-to-date.
As such, the privacy policy might need to be reviewed to include the provision of usage, storage, sharing and publication of crowdsensed data in the policy.

\paragraph{Use or disclosure} An APP entity may collect personal information about an individual for a specific purpose.
Such information must not be used or disclosed for a different purpose without individual’s consent.
In case of exceptional circumstances outlined in APP, the APP entity must take reasonable steps to ensure that the information is de-identified before disclosure.

Smart cities need to collect data from a plethora of sensors, transport the data over the network to servers in the cloud for storage and processing, and use the data to make decision about the city's infrastructure and services in a smarter way.
Although the use of crowdsensed data may not be an issue when developing smart city applications, this system involves several risks for the city and their citizens.

\subsection{Privacy and security risks}
\label{SS:privacy-risks}

Due to involvement of personal information, crowdsensing can inflict three types of risks: 
(1) risk to crowdsensing participant devices, 
(2) risk to the owners of the devices, and,
(3) risk to the crowdsensing data and processing servers.
Firstly, the participating devices bear security risks as these devices could be attacked to enable leakage of sensitive information.
For example, \citet{ronen:2017iot} identified security vulnerabilities of the Philips Hue Smart Light Bulb that could be used to attack these devices.
In addition, the crowdsensing network could be spoofed to collect personal information.
Secondly, the risk may not be limited to devices. It could be extended to the owners of the devices.
Malicious or semi-honest entities in a crowdsensing application could access personal information and use the information to cause harm to the user~\cite{Pournajaf:2014wd, Pournajaf:2016dr}.
For example, the surveillance systems installed at homes could be used to spy on others or the owner itself~\cite{ZhangK:2017}.
Similarly, crowdsensing applications are vulnerable to the Sybil attack~\cite{aisrf:sinai2014} leading to harmful activities against individuals.
Finally, the smart city application server infrastructure bears the risk of being attacked as it contains a lot of data about citizens.
Hence, storage and publication of data, both directly or accidentally, can have several privacy related implications.

\subsection{Privacy implications}
\label{privacy-implications}

Following are some privacy related implications associated with collection and release of data from smart city servers.

\paragraph{Privacy infringement} A business may collect and hold personal information about private citizens when serving them using a crowdsensing based smart city application.
As an APP entity, the business is legally obliged to uphold the privacy law and protect the information.
Publication of data publicly poses the risk of releasing private information if not considered carefully.

\paragraph{Publication of data against law} Publication of data is sometimes prohibited by law due to its sensitivity, or when it infringes someone’s rights or freedoms.
For example, publishing the detailed map of a military site might be prohibited by law.
Yet, when Strava published the heatmap of the fitbit users\footnote{https://labs.strava.com/heatmap/}, it could be used to identify sensitive military locations and their supply routes\footnote{http://www.abc.net.au/news/science/2018-01-29/strava-heat-map-shows-military-bases-and-supply-routes/9369490} publication of which could be deemed against the law.

\paragraph{Trade secret protection infringement} In today's globalisation era, businesses closely interact with each other sharing necessary information with each other.
As such, a business entity may hold information that is sensitive to its business partners and might be considered as trade secrets.
For example, a business might use IoT devices to monitor and manage operations of another business.
If used in crowdsensing application, this data could expose trade secrets of the business involved.

\paragraph{Mosaic effect} Anonymisation is one of the approaches used in privacy protected sharing of data.
Even when the data are anonymised and released publicly, the de-identified data could be combined with other datasets to infer the identify of individuals.
This approach of inferring information by using data from multiple sources, also known as mosaic effect, is of a particular concern.
Even after removing personal data before publishing, data from locations with small populations may still reveal individual identities by implication.
For example, ~\citet{OHara:2011} explained how anonymisation of a subset of data by Google Street View led to identification of the believes and opinions of anonymised entities and made them the subjects of vandalism\footnote{http://www.bbc.com/news/technology-11827862}.

\subsection{Privacy protection mechanisms in crowdsensing}
\label{SS:privacy-protection}

Crowdsensing applications usually have access to either direct personal information or spatio-temporal data that can be used to indirectly infer the personal information.
Several privacy protection mechanisms have been proposed focusing on both direct and indirect access to personal information as elaborated next.
In crowdsensing applications, a single mechanism might not be enough to ensure protection of private information of the participants.
A combination of these mechanisms can be useful~\cite{Blasco:2015ub} as also discussed in the case studies provided in later sections.

\paragraph{Avoidance}
The best way to protect people's private information is not to collect or store it at the first place.
For example, as discussed in Section~\ref{S:casestudy-aadhaar},
if a business needs to authenticate an individual using a third party application, then neither the business nor the third party application should store the information related to the individual's authentication.

\paragraph{Cryptography}
Cryptography techniques are used to ensure secrecy and integrity of data in the presence of an adversary.
Based on the security needs and the threats involved, various cryptographic methods such as \emph{symmetric key cryptography} or \emph{public key cryptography} can be used during transportation and storage of the data.
In addition, a \emph{homomorphic encryption} allows various computations to take place on encrypted data without requiring the data to be decrypted for processing.
From the privacy perspective, these techniques are useful to protect personal information from being leaked during transportation and from storage servers~\cite{Blasco:2015ub}.

\paragraph{Anonymity and Pseudonyms}
When contributing to crowdsensing, the user's identity and location need to be protected.
Anonymisation of users or using pseudonyms can help achieve the protection.
Specifically, the data attributes that are related to user identification can be removed to achieve anonymity.
Similarly, to hide the real identity of users from the cloud infrastructure, trusted third party based pseudonymisers can be used to map the users to pseudonyms~\cite{MartinezBalleste:2013} as well as to separate the location from the data~\cite{To:2017vw}.
Furthermore, methods such as \emph{spatial cloaking} e.g. by using \emph{k-anonymity} improve anonymity by grouping users of relative proximity to each other or by replacing their position with the location of a close point of interest thus a cohort of users share the same location information~\cite{VergaraLaurens:2017hq,To:2017vw}.
Finally, \emph{data aggregation} approaches are used in crowdsensing application to provide summarised data instead of the raw data (e.g.monitoring traffic), thereby removing identifiable information from the data~\cite{Huai:2015,ZhangM:2016,Vakilinia:2016ur}.

\paragraph{Data obfuscation}
Even when the private information of an individual is removed, 
accurate values of the remaining attributes of the record could still be used to identify the individual~\cite{bakken:2004data}.
Data obfuscation can help achieve the user's privacy by transforming private data, e.g.time or location,
in such a way that the adversary cannot infer this data from other data~\cite{bakken:2004data, VergaraLaurens:2017hq, Huning:2017ve}.
For example, a numerical data could be transformed by applying a linear function on the original data to obtain a perturbed data.
Similarly, the original data of a record could be swapped among all the other records to create obfuscation.
In another approach, random noise could be added to the original data to obtain perturbed data.
\emph{Differential privacy} is a obfuscation method in which a randomised function is applied to the original dataset such that
the removal of a single record does not significantly alter the likelihood of an output~\cite{dwork:2008differential}.
It helps protecting privacy of records in a database by adding some randomness to the data~\cite{Joglekar:2017wb, Wang:2016tu, Huai:2015, Sei:bu}.

\paragraph{Access control mechanisms}
In some crowdsensing applications, the private data about the citizen could be collected and stored for legit reasons, e.g.counter terrorism, or providing efficient services to citizen~\cite{UIDAI2011}.
Even when the data is encrypted before storing, the data could still be accessible to all the individuals who have access to the storage systems.
This opens the possibility of insider attack as discussed in section~\ref{S:casestudy-aadhaar}.
Protecting privacy in such cases requires a proper information flow models such as Bell-LaPadula model~\cite{aisrf:bell1976,aisrf:bishop2005}, Lattice model~\cite{aisrf:denning1976} and  Readers-Writers Flow Model (RWFM)~\cite{aisrf:kumar2014} to protect the user's information.

\section{Case Study: Privacy in Crowdsensing for Disaster Management}
\label{S:casestudy1}

A disaster can come in many forms causing loss of lives and severely affecting the economy~\cite{aisrf:miller2006}.
The United Nations International Strategy for Disaster Reduction (UN/ISDR)~\cite{aisrf:unisdr2017} defines disaster as 
a serious disruption of the functioning of a community or a society, at any scale, frequency or onset, due to hazardous events leading to impacting human, material, economic and environmental losses.
The source of disaster can be natural, anthropogenic or both.
Natural disasters are associated with natural processes and phenomena such as hurricane, tsunami and earthquake, while anthropogenic disasters are predominantly induced by human activities, e.g.civil wars.

According to the Emergency Events Database (EM-DAT) figures~\cite{aisrf:cred2016}, a total of 346 natural disaster events occurred globally in 2015 alone.
These events caused 22,773 casualties leaving 98.6 million people affected.
The economic cost of these events was a massive US\$66.5 billion.
Therefore, considering disaster management as a part of developing smart cities is very important.

Information is crucial before, during and after the event of a disaster.
Information and communication technologies (ICTs) have already been used to support the disaster risk management activities~\cite{aisrf:asian2011}.
For example, computer modelling are used to forecast natural disaster warning such as the probability of the occurrence of flood and fire, and the path of a hurricane.
During the disaster event, timely acquisition and processing of data and extraction of accurate information plays a crucial role in providing situation awareness that helps carry on an appropriate disaster response.
Since the people who are in the disaster area including those affected by the disaster accurately know the situation on the ground, they can be the primary source of data.
As such, crowdsensing the situation awareness during disaster is a perfectly sensible approach.

\begin{figure}[t]
\includegraphics[width=\textwidth]{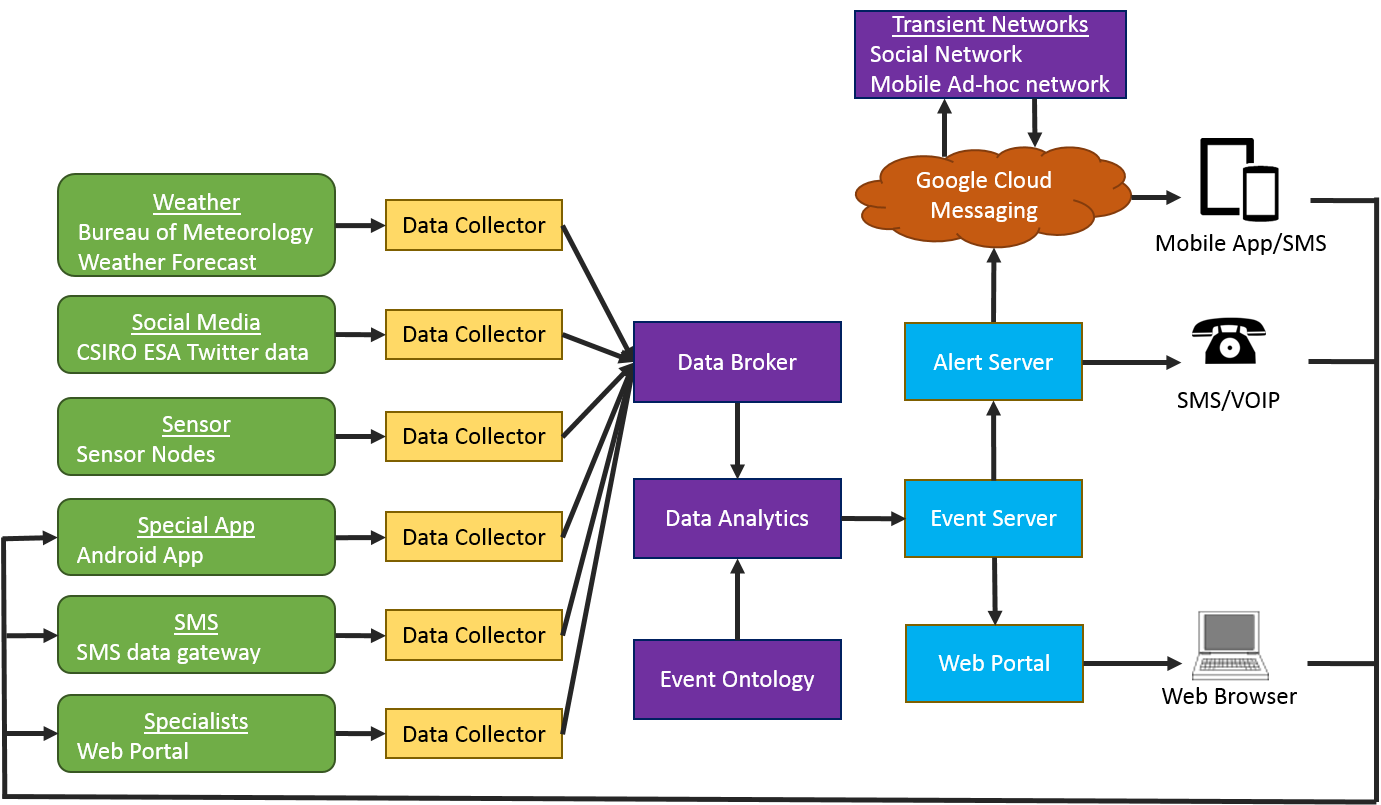}
\caption{A system architecture for disaster management}
\label{fig:architecture}
\end{figure}

We developed a prototype of the situation awareness system using the crowdsensed data from sensors, specialised applications including mobile/web applications, and social media.
Figure~\ref{fig:architecture} demonstrates different crowdsensors required to assess and manage the situation in a natural disaster situation.
In order to understand our approach of crowdsensing in disaster management, let's first consider a disaster scenario.

\subsection{A motivating disaster scenario} 
In Australia, the Crisis Coordination Centre (CCC) is responsible for large-scale disaster management.
The CCC is a round-the-clock all-hazards management facility that provides security, counter terrorism, and the monitoring and reporting of natural disasters and other emergencies.
The CCC has policies and procedures in place for the actions to be undertaken during disaster events.
Part of its remit is the analysis of data from multiple sources, aka data fusion, to understand the scope and the impact of a disaster event.

Imagine that the weather forecast in a certain area predicts heavy rainfall which can possibly cause flooding in the area.
After receiving information from Bureau of Meteorology, an Australian government agency responsible for monitoring weather, climate and water\footnote{http://www.bom.gov.au/},
the CCC creates a transient social network to provide targeted information to the people in the area.
Telstra, a telecommunication provider, provides a list of landline phone numbers installed in the area.
Similarly, all the mobile service providers provide a list of mobile phones that are roaming around in the area.
This information is used to contact people in the area and encourage them to use mobile app, register their phones to receive emergency situation information and enable ad-hoc networking when necessary.
Besides the mobile app, hotlines and SMS services are provided for people to contact emergency services.
John has followed the instructions and is emergency ready.

As expected, severe flooding has just occurred.
John and other people are now providing information to CCC about the situation around them.
John is diabetic and has just fainted.
The app installed in John's phone is sending alert to his doctor about his deteriorating blood sugar levels.
The doctor has informed the authorities, who are now on their way to help him.
People are in touch with their families and friends using social media such as Twitter and Facebook providing information about their situation.
The flood has also caused electricity outage in some areas while the telecommunication network is affected in another area.
Being connected to transient network over ad-hoc network of mobile phones, people in those outage areas are still connected to the rest of the world.
Finally, sensors that are monitoring weather, water quality, air quality etc.are providing crucial information that can affect the health and well-being of people living in the area.

At CCC, data is being processed to gather intelligence about the ground situation.The data is coming from sensors installed in various locations; pictures about the situations that are posted in social media; and the needs at different locations that are posted in CCC application by the ground volunteers.
The information is helping them to make crucial decisions.

\subsection{Crowdsensing in disasters}

In the above scenario, CCC is a specialised entity whose goal is to mitigate the effect of a disaster.
It needs to sense the \emph{situation awareness} as the phenomena of interest.
CCC does not own any infrastructure in the disaster zone.
Therefore, it needs to \emph{sense} the situation awareness from sources that are owned typically by other entities including but not limited to general public.
Hence, it is an exemplary use case of crowdsensing.

Crowdsensing the situation awareness can be achieved by using three types of data sources:
1.sensors, 
2.specialised applications, and, 
3.social networks.
In the above disaster scenario, rain precipitation sensor installed at BOM weather stations as well as at homes of individuals can provide information about the amount of precipitation.
BOM also uses the data to forecast weather pattern of coming days and generate warnings.
These information can be used to predict the level of flood in an area and plan for disaster mitigation and recovery.
People can use specialised applications such as the triple zero app\footnote{http://emergencyapp.triplezero.gov.au/} to alert emergency services about the emergency needs.
Use of SMS in emergency situation is common in many countries.
Similarly social networks can sometimes be turned into specialised applications in disaster situations.
For example, Facebook can be converted to a crisis response app in case of emergencies\footnote{https://www.facebook.com/about/crisisresponse/}.
Similarly, Twitter can be used to detect a disaster as well as assess ground level information about the situation\footnote{https://esa.csiro.au/}.
In all the above examples, the situation awareness is opportunistically sensed by using data generated by several independent entities.

In situations arising out of disaster management, unstable, unreliable ad hoc channels could become the only means of communication and information gathering.
The characteristics of such communication can be listed as follows:
\begin{itemize}
\item The nature of the disaster related communication, in the context of smart city, is information gathering.
\item Information on the ground zero is generated either through crowd sensing such as real-time twitter, or submission of information by mobile digital volunteers to a crowdsourcing platform.
\item Mostly the information flow is carried out through wireless networks which provide unstable, intermittent connectivity and through open public networks.
\item The other major sources of information are wireless sensor and actuator installations specially established for detecting events such as mudslide, flood, storm, fire, volcanic or seismic activities. 
\end{itemize}

Thus the network paths have to overcome a whole range of network interoperability and connectivity issues. In disaster situation,
we specifically explored the uses of delay tolerant network to create transient social network~\cite{aisrf:bhatnagar2016}.
We further assessed the privacy aspects in disaster management, particularly when using such applications as discussed next.

\subsection{Privacy in disaster management}

Leaving the technical problems like connectivity and internetwork operability aside, the privacy issue in information gathering and use are minimal.
In this case, there are two possible approaches to data gathering: 
(i) information is shared voluntarily by user of the device, 
(ii) involuntarily from embedded device sensors.
In both cases, location information and other individual details are important and may still be embedded inside the collected data.
Such information may involuntarily leak the user's privacy as it may provide enough information about the movement of the user.
Even digital volunteers may expect anonymization of information they would share through the first approach.

In our approach, we considered security and privacy in both mobile phone apps and sensors based IoT systems.
Sybil attack is a possible attack in a smart phone based disaster management application~\cite{aisrf:sinai2014}.
The Sybil attack is an attack wherein a reputation system is subverted by forging identities in peer-to-peer networks.
The lack of identity in such networks enables the bots and malicious entities to simulate fake GPS report to influence social navigation systems.
The Sybil attack is more critical in a disaster situation where people are willing to help the distressed person.
The vulnerability could be misused to compromise people's safety.
For example, the malicious user can simulate a fake disaster alarm to motivate a good Samaritan to come for help in a lonely place and cause harm.
The attacker can also divert attention of rescue team from a real disaster.
Using appropriate cryptography combined with an appropriate access control model could help disseminate information while protecting privacy in a crowdsensing application.

People try to do their best to communicate with others when they are in a distressed situation.
In this situation, one challenge is to disseminate information while controlling the access to the information. 
Proper information flow models such as Bell-LaPadula model~\cite{aisrf:bell1976,aisrf:bishop2005}, Lattice model~\cite{aisrf:denning1976} and  Readers-Writers Flow Model (RWFM)~\cite{aisrf:kumar2014} can be used to protect the user's information.
Combined with proper access control mechanisms, those information flow model can be used to guarantee that the information flow follows the required privacy rules and does not leak any critical information to the adversary.
For example, in RWFM, the sender can control the readers of a message by specifying their names in the readers list.

Another hard challenge in this situation is to enable end-to-end security and privacy in processing big data streams emitted by geographically distributed mobile phones and sensors.
We have investigated and proposed a number of techniques (refer to~\cite{aisrf:puthal2015a,aisrf:puthal2015,aisrf:kumar2016,aisrf:puthal2017,aisrf:puthal2017a} for details).
Applications in risk-critical domains such as disaster management need near-real-time stream data processing in large-scale sensor networks.
We introduced a Data Stream Manager (DSM) to perform security verification just before stream processing engine.
DSM works by removing the modified data packets and supplying only original data back to the steam processing engine for evaluation.
Furthermore, we proposed a Dynamic Key-Length-Based Security Framework (DLSeF) based on a shared key derived from synchronized prime numbers;
the key is dynamically updated at short intervals to thwart potential attacks~\cite{aisrf:puthal2015}.
DLSeF has been designed based on symmetric key cryptography and dynamic key length to provide more efficient security verification of big sensing data streams.
Furthermore, to secure big sensing data streams we have also proposed Selective Encryption (SEEN) method that satisfies the desired multiple levels of confidentiality and data integrity~\cite{aisrf:puthal2017}.

In this way, we ensured that the data is encrypted when flowing across the mobile ad-hoc network and when stored in a database while the read-write flow model is used to ensure access to the data is controlled appropriately.

\section{Case Study: Citizen Privacy in Aadhaar}
\label{S:casestudy-aadhaar}

An entity is responsible for maintaining security of private information of the users of its infrastructure and services.
These responsibilities are broadly defined in law as discussed earlier in Section~\ref{S:privacy}, specifically in Section~\ref{SS:privacy-law-australia} in Australian context.
Governments are the entities who often need to uniquely identify their citizens in order to efficiently and seamlessly provide necessary services.
For example, the US government issues social security number (SSN)\footnote{https://www.ssa.gov/ssnumber/} to provide and monitor social securities to its residents. The SSN of an individual is considered as sensitive and secret.
Australian government requires individuals to obtain tax file number (TFN)\footnote{https://www.ato.gov.au/Individuals/Tax-file-number/} for tax purposes, medicare card for medical benefits and centerlink number for social services\footnote{https://www.humanservices.gov.au/}.
An attempt of building an identity platform for the UK was dropped at the intervention of its Parliament\footnote{https://www.gov.uk/identitycards}.

Recently, the Government of India developed an identity platform called Aadhaar\footnote{https://uidai.gov.in/en}. 
Aadhaar is a biometrics based identity database and authentication system.
Since today's mobile phones and personal devices are capable of collecting biometric information of their users, they could be used to connect to Aadhaar to ascertain the users' identities for crowdsensing applications in smart cities. 
As such, it 
offers an excellent case study and 
shows the need for an integrated working of technologies, processes and the law to realize the needed privacy along with the bearers of trust.

\subsection{About Aadhaar}
The government of India has set up an exclusive agency called Unique Identification Authority of India (UIDAI) to build an identity platform called Aadhaar for people residing in India including non-citizens.
Aadhaar has a huge central database that includes demographic and biometrics information of over 1.2 billion individuals. 
It has a significant amount of person centric information including multiple biometrics such as photographs, finger prints and images of iris from which a person can be uniquely identified. 
One of the main reasons for creating Aadhaar as a huge Social Identification System (SIS) had been to prevent massive leakages and large scale fraudulent transactions in implementation of targeted delivery of subsidies for the poor.

\begin{figure}[t]
\begin{center}
\includegraphics[width=\textwidth,scale=0.7]{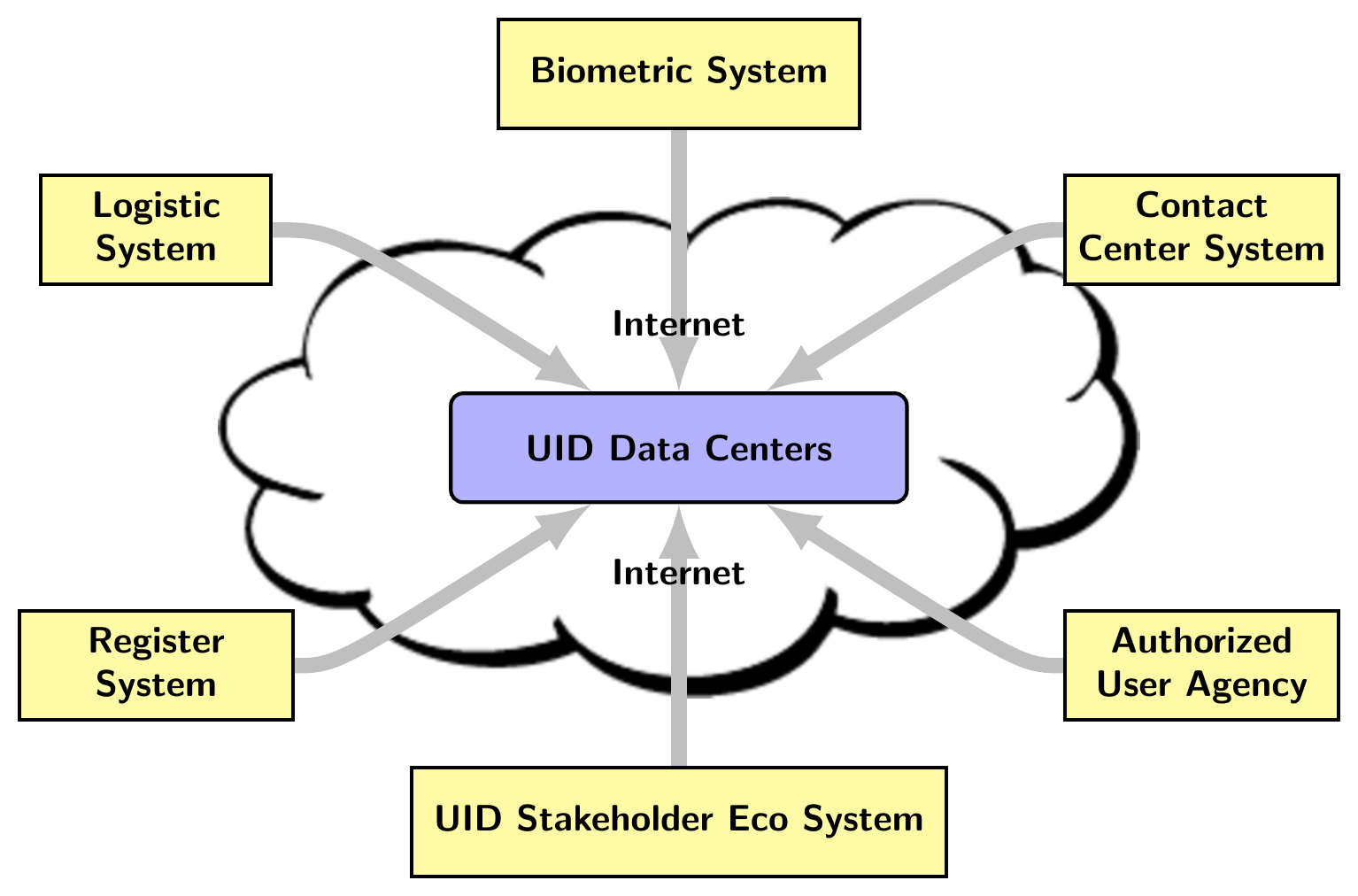}
\end{center}
\caption{A top level view of Aadhaar operation ecosystem.}
\label{fig:aadhaarEco}
\end{figure}

The government of India intends to use this platform not only to provide and monitor the delivery of social services
but also to use it as an identity verification system for wide range of purposes. 
It already provides an authentication service to Authentication User Agencies (AUAs) such as civil supplies, insurance companies and banks to affirm the identities of individuals based on their biometrics and Aadhaar number. A top level view of Aadhaar operating ecosystem~\cite{zelazny2012} is provided in Fig.~\ref{fig:aadhaarEco}.

UIDAI claims that its implementation of SIS system ensures that AUAs cannot infringe on the user's privacy.
UIDAI dismisses any scope of massive scale surveillance as being alleged by the rights activists.
In its defence, UIDAI technology support groups reveal its technology stack having multiple layers of autonomous entities.
There is a Central IDentities Repository (CIDR) which provides Identity service, enrolment packets, authentication log, variety of meta data as well as transactional logs.
Identification service requests are channelised through Authentication Service Agencies (ASAs) working as intermediary or third party verification agencies, between CIDR and AUA.
ASAs must have prior registrations in order to avail CIDR services.

Note that several apps including the ones by Government agencies have been built on the Aadhaar platform, and are in wide use.
One can gauge the possibilities of privacy leaks and compromises on personal information through them. 
Before we delve into the privacy issues with Aadhaar, let us present the necessary background information first.

\subsection{Authentication and Authorisation}
\label{SS:authentication-authorisation}

In 2007, the OECD developed recommendation on electronic authentication for its member countries~\cite{oecd:authentication2007}.
India has a working relationship with OECD as a non-member country.
According to the recommendation, authentication is \emph{``A function for establishing the validity and assurance of a
claimed identity of a user, device or another entity in an
information or communications system''}.
One of the principles advocated in the OECD guidelines for authentication/authorization is,  
{\em ``Do not include \textit{authorisation} (which is a separate but a related process that
refers to verifying the person's or organisation's authority to conduct specified
transactions).
Typically, decisions concerning authorisation are the purview
of the relying party (i.e., the entity or person that is relying on the identity
assertion to make the authorisation decision).''}

As Aadhaar has a huge central biometric database being used for authentication (or identification), it is necessary to look at the issues of security/privacy in such context.

\subsection{Biometrics and Privacy}
\paragraph{Biometrics is not secret} While biometrics such as fingerprints, iris scans and facial images are \emph{private}, they are not \emph{secrets}.
We leave a copy of our fingerprints on almost everything we touch.
A modern smartphone is highly capable of taking high resolution pictures of our faces from which the iris biometrics can be extracted.
Thus, unless we spend our life wearing gloves and shades,
there is no hope that our biometrics can be kept secret.
Just as our names, our biometrics are available to the people we encounter in our daily lives. Arguably, our names could be treated as being more secret than our biometrics.

Technically anyone can create a database of biometrics of people in public.
However, such a possibility can be discarded for the following reasons. 
Firstly, 
unlike one's name, other people have no use for one's biometrics in normal daily life and hence, people don't pay attention to them.
For this reason, people may mistakenly believe that their biometrics are secret.
Secondly, although people's biometrics is publicly available, there is no public database from which biometrics can be freely downloaded. Such a database would be deemed illegal in many countries.

\paragraph{Biometrics Matching}
Biometric matching techniques gives probabilistic, and not deterministic, answers.
In other words, the biometrics scanner scores a match on a scale of 0\% to 100\%.
It cannot give a straightforward \emph{yes} or \emph{no} answer.
In Aadhaar, the UIDAI technical support group claims~\cite{UIDAI2011} that the False Positive Identification Rate (FPIR) of biometrics matching is as low as 0.0025\% for every 1:N searches.
Also, UIDAI has the capacity to scale up to one million enrolments per day.
Therefore, in the full scale enrolment process, just about 25 false positive matches may need to be manually judged.
At this rate, the false positives will not create major problems even if the gallery continuously keeps on growing.

Since \emph{biometric} is \emph{at most private and never a secret},
using biometric for authentication has a few issues as discussed below.

\subsection{Biometrics and Authentication}
The current size of the Aadhaar biometrics gallery is over 1.2 billion records.
When using biometrics together with other attributes such as Aadhaar number or name, the chances of getting both of them matched together are still very low at 0.0025\% false positive rate. 
In other words, it is comparable to using \emph{username} and \emph{password} for authentication where the username is a unique identity and the password is associated to the username but is secret.
Now, if the biometrics alone is used, then this scenario will be similar to using password alone  for authentication/authorisation, and even the password is not secrete as explained above. 
Moreover, if the authority of authentication is given to a scanner/server connected to UIDIA,
then each biometrics scan will produce over 30,000 matches and hence this process will fail to adequately authenticate an individual. 
In addition, if an individual can obtain the Aadhaar number of any of the 30,000 other individuals whose biometrics identities are similar to its own biometrics identities, then the individual can flawlessly acquire the other individual's identity and be able to access services including social as well as financial services.
Even if an individual's biometrics do not match with another individual's biometrics, the individual can still break the authentication system by cloning the other individual's identity as demonstrated by BAIDU Security Lab\footnote{http://sgcsc.sg/doc/Camp2/WS-06.pdf}.
Finally, the UIDAI base automatic authentication system could potentially lead to {\em Denial of Service} as all the records need to be compared for each record search. 
Thus, leaving authority of authentication/identity to electronic devices and the server is at risk.

\subsection{Privacy Pitfalls of Authentication}
\label{SS:privacy-pitfalls}
According to~\citet{aisrf:schneider}, a textbook material from a yet to be published book 
on Cybersecurity by Prof. Fred Schneider, a leading security expert from Cornell University who has championed several Cybersecurity guidelines for US, authentication when undertaken injudiciously can lead to privacy violations for the following reasons.

Firstly, in authenticating somebody, you learn its identity and thus,
you also learn an associated set of attributes some of which could be considered personal information.
Thus, authentication could lead to the revelation of personal information.

Secondly, threat to privacy arises when authentication is used to validate participants in some action.
It is possible that participation is deemed private (for instance some certain medical purchases or medical procedures); 
thus the side effect of authentication is to associate personal information with an identity.
This problem is compounded when the same identifier is used to authenticate an individual in connection to multiple actions
leading to the capabilities of third parties to be able to 
connect seemingly unrelated actions with a single individual and then 
make inferences about associated additional attributes of that individual.

Thirdly, in a sense, requirement of authentication implicitly institutes a form of authorization.
Thus, the prospect of undergoing authentication inhibits people from engaging in activities they fear could be misconstrued, deemed inappropriate, or lead to retribution.
The concern here is not that there is an erosion of basic freedoms when authentication is required but that this erosion is inadvertent.
The concern is that the policy
-- {\em not side-effects of a system's construction} -- should be what dictates who may engage in what activities, and authorization -- {\em not authentication }-- mechanisms
should be what implements such policy.

Finally, an authentication system collects and possibly stores information for subsequent use.
\textit{Note that information collected without consent should never be allowed to be abused (that includes linking etc.)}.

Consequently, widespread deployment of authentication mechanisms increases plausible privacy violations in three ways: 
(i)  personal information could
be abused by the agency collecting it, 
(ii) stored personal information could be stolen, and 
(iii) having  personal information further increases the risk
of inferences by linking shared identities or other shared attributes.

\subsection{Authentication Guidelines}
Some of the broad guidelines that exist in the context of instituting e-authentications including crowdsensing are~\citep{aisrf:schneider}:

\paragraph{Seek Consent}
The entity performing authentication should seek consent from users about what information will be stored and what will be revealed to the authenticating entities.
In this way, people become aware that they are relinquishing some control over the confidentiality of personal information to that identity.

\paragraph{Select Minimal Identity} 
Authentication should be performed only against identities that embody the smallest set of attributes needed for the task. 
This reduces unnecessary exposure of privacy related attributes to the system.

\paragraph{Limit Storage} 
The information about authenticated identities should not be saved unless there is a clear need. 
Moreover, the information should be deleted once it is no longer needed.
This reduces the chances that the saved identity information can subsequently be re-targeted for uses not implied by the consent of the user that allowed its collection.

\paragraph{Avoid Linking} 
A single, shared attribute allows linking the
identities that contain this attribute, 
and that could violate privacy by revealing attributes comprising one identity to those who learn the other identity.
Thus, it is important to eschew including the same unique attribute (e.g., identifier) in different identities.

Now, let us turn to UIDAI claims regarding Aadhaar.

\subsection{Issues of Privacy Leaks with Aadhaar}
Looking at the issues of biometric databases as well the guidelines for avoiding privacy violations discussed above, 
it is needless to say that preventing privacy violations in the cyberspace for the largest database of biometrics in the world 
will be a grand challenge.

\paragraph{User Expectation}
People do provide their biometrics for varieties of purposes including visa for foreign countries. Therefore, the main question here is, \emph{Is Aadhaar at least as safe as those systems?} 

Two important characteristic differences from Aadhaar operation in the context of visa issuance are:
\begin{enumerate}
\item The biometric database is exclusively used for the purpose of issuance of visa with a limited storage as highlighted above against the passport cum visa number of the individual.
\item The authority of authentication is with the immigration officer and not the server of the database.
\end{enumerate}
Thus, one expects a trusted process that involves minimal invasion of privacy with consent.

\paragraph{Insider Attacks}
\label{SS:insider-attack}
As highlighted already, privacy characterization is dependent on the law of the land.
Keeping this in mind, there has been an important case (or a bunch of cases) currently at the Supreme Court of India for adjudication about the UIDAI claims and the counter claims made by the activists spearheading the assurance of privacy preserving implementation.
This will have a long-standing bearing on the interpretation of privacy law in the context of technology.
The basic premise in all above mentioned privacy issues mainly stems from the insider attacks.

From a user perspective, the insider attack on privacy issues can be classified as:
\begin{enumerate}
\item Protection of privacy against registration authority.
\item Protection of privacy against access network operators.
\item Protection  of privacy against attacks using a set of correlated queries by a set colluding agencies.
\end{enumerate}

In the context of the first type of insider attack,
in January 2018, a leak of Aadhaar numbers was reported by a journalist who paid just Rs 500 (equivalent to $\approx$ US \$8) to access personal data from an enrolment centre.
Even though this event may look insignificant,
there are possible dire consequences from such incidents.
Specifically, the Aadhaar system has a false positive rate of 0.0025\% for a biometric match.
Arguably, for each individual in the database, there are possibly 30,000 individuals whose biometrics identities match with the individual in a database of 1.2 billion individuals.
Consequently, if one can collude with the insiders to obtain Aadhaar number of someone else whose biometric profiles matches with one's own biometric profile, 
it will be able to use the obtained identity for various purposes including financial gains or even criminal activities. 

Although, one can argue that the processes can be strengthened to avoid such possibilities, 
it clearly demonstrates that a group of colluding insiders can compromise a surveillance system and breach both privacy and security.

In the contexts of other two types of attacks, these insider attacks are technical in nature.
Particularly, the attack of type 3 appears to be technically extremely difficult.
Fortunately, creating such attacks need an active and cohesive collaboration across agencies, which is difficult to accomplish and prone to leakage.
A multi-authority strong attribute based encryption framework~\cite{Chase:2009,Li2013} could possibly alleviate these problems.
The kernel level support for role based decryption rights~\cite{kuhn2000,sandhu2000,bertino2001} could also enhance the trustworthiness of data.  

\subsection{Security of Aadhaar}
\label{SS:aadhaar-security}
The CEO of UIDAI on 22 March 2018 presented to the Supreme Court of India, the physical security, details of 2048 bit RSA encryption\footnote{For an assessment of 2048 encryption, refer \cite{aisrf:rsa2048}.}, 
security reviews conducted etc. (refer blog \cite{aisrf:vombatkere} for the technical points presented by the CEO of Aadhaar).
In biometric databases of the magnitude of Aadhaar, 
it is important to ensure that long-term keys do not compromise past session keys.
It would be nice if properties such as \textit{forward secrecy (FS)} are implemented in Aadhaar. 
Forward secrecy protects past sessions against future compromises of secret keys or passwords.
A perfect forward secrecy (PFS) assures compromises of long-term keys in the communication protocols do not compromise past session keys.
Whether this property is satisfied in the process provided by Aadhaar \cite{aisrf:aadhaarprotocol} needs a thorough analysis.

As highlighted above, in the context of Aadhaar, the same database is used for varieties of purposes and hence, linking of several authentications exist on the same identifier.
Thus, it is not easy to prove that there is no violation of privacy.
Even the claim of \textit{no storage} by Aadhaar needs careful analysis in the context of using profiles to track identities as revealed in the infamous leakage on Facebook.
The experiment done~\cite{aisrf:vtprks} shows the privacy leakage from a wide spectrum of angles.

As per the UIDAI website, Aadhaar can be used both for identity 
(one can download e-Aadhaar on one's smartphone that will serve as an identity)
as well as for authentication.
Thus, the process has to make it clear the relationship of the biometric usage for accessing the database.



\subsection{Issues in Anonymization and Virtual IDs}
\label{SS:issues-virtual-ids}
Recently, UIDAI has announced the use of Virtual IDs (creatable by the user) would be used for authentication rather than the persons' Aadhaar number.
While it appeals generally, one need to be clear about the possibilities of compromise\footnote{In the era of social network profiling compounded with privacy breaches like the recent (March 2018) Facebook data breach} 
of ones' Aadhaar number, virtual numbers still looks risky 
(one need to answer the question whether Aadhaar number is a private or a secret number -- if it is used as an identity, certainly it cannot be a secret number).
Note that biometric access control is used in a controlled environments like accessing server rooms, airport gates etc., and not in a public environments as the biometric can be cloned as highlighted already.

In general, privacy concerns in crowdsensing arise from the requirements of disclosure of personal information such as IMEI number of mobile, SIM ID, Phone number, IP address, location, cell tracking, and other subscriber related information.
From such information, one can easily extract the home or the office address of the person.
Therefore anonymization\footnote{Recently, Facebook Inc., admitted that data on as many as 87 million people, most of them in the U.S., may have been improperly shared with research firm Cambridge Analytica. The challenge is to understand, how good is the anonymization and how much profiling can it withstand will remain a question. As in the case of authentication, it may inhibit people sharing data (even under anonymization) and  engaging in activities they fear
could be misconstrued or exploitable.} of data appears to be the only way to encourage individual to share \textit{the} data.
However, complete anonymization could create problem of malicious individuals sharing spurious and concocted data.
So, there is a need to balance between two conflicting requirements, the anonymization and the trustworthiness of data, in order to establish an effective privacy preserving crowdsensing mechanism for surveillance/smartness.

Typically, servers managed by a trusted third party (like it is done in authenticating credit card transactions, DNS etc.,) are made available for gathering data through crowdsensing.
Alternatively, an individual can upload videos or images on a specified third party server.
This server removes all traces of tracking information (e.g., personal IDs, phone numbers, GPS data, car license numbers) before sending an event related data to the security surveillance and control system.
The server itself should be accessible only through multiple layers of strong cryptographic frameworks.
Most crowdsensing system incorporate reputations of participants tailored to specific surveillance system~\cite{Dua:2009,huang2010,huang2012,christin2013}.
Incognisense~\cite{christin2013} framework proposed use two basic features: 
(i) using multiple virtual IDs with dynamic or periodic changes, and, 
(ii) mapping exact reputation values to a range, where a range is mapped to a reputation group.
Mapping reputation to range and the use of dynamic virtual ID completely masks the ID of individual participants.
However, this add extra overhead in the management of crowdsensed data either in terms of complex algorithm (for dynamic assignment of pseudo names) or in terms of maintaining redundancy on participant information~\cite{he2015}.

\section{Conclusion}
In the age of digital technologies where a personal phone is capable of collecting, storing and transmitting personal data even without the person's conscious knowledge,
privacy issues have become highly complicated and extremely important.
Particularly, in case of crowdsensing 
where an individual's smartphone and other personal devices embedded with modern sensors are involved in collecting and transmitting information,
privacy becomes a real issue.

In this chapter, we explored crowdsensing in smart city applications and its privacy implications.
Firstly, we presented various definitions related to the term crowdsensing.
We demonstrated that the current definition of mobile crowdsensing that is used interchangeably with crowdsensing is not appropriate since crowdsensing could be done by using non-mobile sensors as well as by using people as social sensors.
Secondly, we explored various aspects of privacy, including legal definitions, obligations to crowdsensing service providers, 
risks, implications and possible solutions.
Thirdly, we presented two case studies relevant to crowdsensing. 
In the first case study, we demonstrated how crowdsensing can be used in disaster management.
Although the privacy concerns during a disaster event could be considered minimal, 
the fact that the data collected during the disaster can remain available long after the event is a real concern.
As such, the privacy issues in this case study are managed by using cryptography and appropriate access control mechanisms.
In the second case study, we discussed how citizen information can be collected by the governments and how the citizen's private data can be prone to privacy risks.
We discussed the possibility that the authorities could compromise a system to get access to people's personal information is not unreal.
In all these cases, there is no single solution that can solve all the privacy problems in these crowdsensing applications.
However, a combination of possible solutions can be used to achieve the maximum possible protection of the private information.

On one hand, people contributing to crowdsensing applications to make their cities smarter need to easily and clearly understand how their private data are going to be used.
Therefore, further research is required to make the understanding of privacy issues easier and more controllable.
On the other hand, the entities working with crowdsensing data containing people's private information need to have adequate knowledge of their privacy obligations, implications, and appropriate protection mechanisms. 
As the attempt to use crowdsenseing data for making smart decisions in smart cities will continue, 
on going research will be required to identify and address privacy issues in those applications.

\section*{Acknowledgements}
This research is funded by Australia India Strategic Grant AISRF-08140.

\section*{References}
\bibliographystyle{model1-num-names}
\bibliography{references}

\begin{thebibliography}{72}
\expandafter\ifx\csname natexlab\endcsname\relax\def\natexlab#1{#1}\fi
\providecommand{\bibinfo}[2]{#2}
\ifx\xfnm\relax \def\xfnm[#1]{\unskip,\space#1}\fi
\bibitem[{Giffinger et~al.(2007)Giffinger, Fertner, Kramar, Kalasek,
  Pichler-Milanović, and Meijers}]{Giffinger2007}
\bibinfo{author}{R.~Giffinger}, \bibinfo{author}{C.~Fertner},
  \bibinfo{author}{H.~Kramar}, \bibinfo{author}{R.~Kalasek},
  \bibinfo{author}{N.~Pichler-Milanović}, \bibinfo{author}{E.~Meijers},
  \bibinfo{title}{{Smart cities - Ranking of European medium-sized cities}},
  \bibinfo{year}{2007}.
\bibitem[{Caragliu et~al.(2011)Caragliu, Del~Bo, and
  Nijkamp}]{caragliu2011smart}
\bibinfo{author}{A.~Caragliu}, \bibinfo{author}{C.~Del~Bo},
  \bibinfo{author}{P.~Nijkamp},
\newblock \bibinfo{title}{Smart cities in europe},
\newblock \bibinfo{journal}{Journal of urban technology} \bibinfo{volume}{18}
  (\bibinfo{year}{2011}) \bibinfo{pages}{65--82}.
\bibitem[{Ganti et~al.(2011)Ganti, Ye, and Lei}]{Ganti2011}
\bibinfo{author}{R.~Ganti}, \bibinfo{author}{F.~Ye}, \bibinfo{author}{H.~Lei},
\newblock \bibinfo{title}{{Mobile crowdsensing: current state and future
  challenges}},
\newblock \bibinfo{journal}{IEEE Communications Magazine} \bibinfo{volume}{49}
  (\bibinfo{year}{2011}) \bibinfo{pages}{32--39}.
\bibitem[{Lefort et~al.(2011)Lefort, Henson, Taylor, Barnaghi, Compton, Corcho,
  Garcia-Castro, Graybeal, Herzog, Janowicz et~al.}]{lefort2011semantic}
\bibinfo{author}{L.~Lefort}, \bibinfo{author}{C.~Henson},
  \bibinfo{author}{K.~Taylor}, \bibinfo{author}{P.~Barnaghi},
  \bibinfo{author}{M.~Compton}, \bibinfo{author}{O.~Corcho},
  \bibinfo{author}{R.~Garcia-Castro}, \bibinfo{author}{J.~Graybeal},
  \bibinfo{author}{A.~Herzog}, \bibinfo{author}{K.~Janowicz}, et~al.,
\newblock \bibinfo{title}{Semantic sensor network xg final report},
\newblock \bibinfo{journal}{W3C Incubator Group Report} \bibinfo{volume}{28}
  (\bibinfo{year}{2011}).
\bibitem[{Noble et~al.(1980)Noble, Gill, and Bary}]{aisrf:noble1980}
\bibinfo{author}{I.~R. Noble}, \bibinfo{author}{A.~M. Gill},
  \bibinfo{author}{G.~A.~V. Bary},
\newblock \bibinfo{title}{Mcarthur's fire-danger meters expressed as
  equations},
\newblock \bibinfo{journal}{Austral Ecology} \bibinfo{volume}{5}
  (\bibinfo{year}{1980}) \bibinfo{pages}{201--203}.
\bibitem[{Roitman et~al.(2012)Roitman, Mamou, Mehta, Satt, and
  Subramaniam}]{Roitman:2012kw}
\bibinfo{author}{H.~Roitman}, \bibinfo{author}{J.~Mamou},
  \bibinfo{author}{S.~Mehta}, \bibinfo{author}{A.~Satt}, \bibinfo{author}{L.~V.
  Subramaniam}, \bibinfo{title}{{Harnessing the crowds for smart city
  sensing}}, \bibinfo{publisher}{ACM}, \bibinfo{address}{New York, New York,
  USA}, \bibinfo{year}{2012}.
\bibitem[{Power et~al.(2013)Power, Robinson, and Ratcliffe}]{aisrf:power}
\bibinfo{author}{R.~Power}, \bibinfo{author}{B.~Robinson},
  \bibinfo{author}{D.~Ratcliffe},
\newblock \bibinfo{title}{Finding fires with twitter},
\newblock in: \bibinfo{booktitle}{Australasian Language Technology Association
  Workshop}, volume~\bibinfo{volume}{80}, pp. \bibinfo{pages}{80--89}.
\bibitem[{Yin et~al.(2012)Yin, Lampert, Cameron, Robinson, and
  Power}]{aisrf:yin2012}
\bibinfo{author}{J.~Yin}, \bibinfo{author}{A.~Lampert},
  \bibinfo{author}{M.~Cameron}, \bibinfo{author}{B.~Robinson},
  \bibinfo{author}{R.~Power},
\newblock \bibinfo{title}{Using social media to enhance emergency situation
  awareness},
\newblock \bibinfo{journal}{IEEE Intelligent Systems} \bibinfo{volume}{6}
  (\bibinfo{year}{2012}) \bibinfo{pages}{52--59}.
\bibitem[{Guo et~al.(2015)Guo, Wang, Yu, Wang, Yen, Huang, and
  Zhou}]{guo2015mobile}
\bibinfo{author}{B.~Guo}, \bibinfo{author}{Z.~Wang}, \bibinfo{author}{Z.~Yu},
  \bibinfo{author}{Y.~Wang}, \bibinfo{author}{N.~Y. Yen},
  \bibinfo{author}{R.~Huang}, \bibinfo{author}{X.~Zhou},
\newblock \bibinfo{title}{Mobile crowd sensing and computing: The review of an
  emerging human-powered sensing paradigm},
\newblock \bibinfo{journal}{ACM Computing Surveys (CSUR)} \bibinfo{volume}{48}
  (\bibinfo{year}{2015}) \bibinfo{pages}{7}.
\bibitem[{Taylor et~al.(2013)Taylor, Griffith, Lefort, Gaire, Compton, Wark,
  Lamb, Falzon, and Trotter}]{taylor2013farming}
\bibinfo{author}{K.~Taylor}, \bibinfo{author}{C.~Griffith},
  \bibinfo{author}{L.~Lefort}, \bibinfo{author}{R.~Gaire},
  \bibinfo{author}{M.~Compton}, \bibinfo{author}{T.~Wark},
  \bibinfo{author}{D.~Lamb}, \bibinfo{author}{G.~Falzon},
  \bibinfo{author}{M.~Trotter},
\newblock \bibinfo{title}{Farming the web of things},
\newblock \bibinfo{journal}{IEEE Intelligent Systems} \bibinfo{volume}{28}
  (\bibinfo{year}{2013}) \bibinfo{pages}{12--19}.
\bibitem[{Karimi et~al.(2015)Karimi, Wang, Metke-Jimenez, Gaire, and
  Paris}]{karimi:2015}
\bibinfo{author}{S.~Karimi}, \bibinfo{author}{C.~Wang},
  \bibinfo{author}{A.~Metke-Jimenez}, \bibinfo{author}{R.~Gaire},
  \bibinfo{author}{C.~Paris},
\newblock \bibinfo{title}{Text and data mining techniques in adverse drug
  reaction detection},
\newblock \bibinfo{journal}{ACM Computing Surveys (CSUR)} \bibinfo{volume}{47}
  (\bibinfo{year}{2015}) \bibinfo{pages}{56}.
\bibitem[{Brabham(2008)}]{Brabham:2008jh}
\bibinfo{author}{D.~C. Brabham},
\newblock \bibinfo{title}{{Crowdsourcing as a Model for Problem Solving}},
\newblock \bibinfo{journal}{Convergence: The International Journal of Research
  into New Media Technologies} \bibinfo{volume}{14} (\bibinfo{year}{2008})
  \bibinfo{pages}{75--90}.
\bibitem[{O'Hara(2011)}]{OHara:2011}
\bibinfo{author}{K.~O'Hara}, \bibinfo{title}{Transparent government, not
  transparent citizens: a report on privacy and transparency for the cabinet
  office}, \bibinfo{year}{2011}.
\bibitem[{Cohen(2013)}]{Cohen:2012}
\bibinfo{author}{J.~E. Cohen},
\newblock \bibinfo{title}{{What Privacy Is For}},
\newblock \bibinfo{journal}{Harvard Law Review} \bibinfo{volume}{126}
  (\bibinfo{year}{2013}).
\bibitem[{Solove(2011)}]{Solove:2011}
\bibinfo{author}{D.~J. Solove},
\newblock \bibinfo{title}{Why privacy matters even if you have 'nothing to
  hide'},
\newblock \bibinfo{journal}{The Chronicle of Higher Education}
  (\bibinfo{year}{2011}).
\bibitem[{Magi(2011)}]{Magi:2011}
\bibinfo{author}{T.~J. Magi},
\newblock \bibinfo{title}{Fourteen reasons privacy matters: A multidisciplinary
  review of scholarly literature},
\newblock \bibinfo{journal}{Library Quarterly} \bibinfo{volume}{81}
  (\bibinfo{year}{2011}) \bibinfo{pages}{187--209}.
\bibitem[{Cilliers and Flowerday(2014)}]{Cilliers:2014}
\bibinfo{author}{L.~Cilliers}, \bibinfo{author}{S.~Flowerday},
\newblock \bibinfo{title}{{Information security in a public safety,
  participatory crowdsourcing smart city project}},
\newblock in: \bibinfo{booktitle}{2014 World Congress on Internet Security
  (WorldCIS)}, \bibinfo{publisher}{IEEE}, \bibinfo{year}{2014}, pp.
  \bibinfo{pages}{36--41}.
\bibitem[{Martinez-Balleste et~al.(2013)Martinez-Balleste, Perez-Martinez, and
  Solanas}]{MartinezBalleste:2013}
\bibinfo{author}{A.~Martinez-Balleste}, \bibinfo{author}{P.~A. Perez-Martinez},
  \bibinfo{author}{A.~Solanas},
\newblock \bibinfo{title}{{The Pursuit of Citizens' Privacy: A Privacy-Aware
  Smart City Is Possible}},
\newblock \bibinfo{journal}{IEEE Communications Magazine} \bibinfo{volume}{51}
  (\bibinfo{year}{2013}) \bibinfo{pages}{136--141}.
\bibitem[{Garcia(2017)}]{Garcia:2017ch}
\bibinfo{author}{D.~Garcia},
\newblock \bibinfo{title}{{Leaking privacy and shadow profiles in online social
  networks}},
\newblock \bibinfo{journal}{Science Advances} \bibinfo{volume}{3}
  (\bibinfo{year}{2017}) \bibinfo{pages}{e1701172}.
\bibitem[{Li et~al.(2017)Li, Jeong, Shin, and Park}]{LiY:2017}
\bibinfo{author}{Y.~Li}, \bibinfo{author}{Y.-S. Jeong}, \bibinfo{author}{B.-S.
  Shin}, \bibinfo{author}{J.~H. Park},
\newblock \bibinfo{title}{{Crowdsensing Multimedia Data: Security and Privacy
  Issues}},
\newblock \bibinfo{journal}{{IEEE Multimedia}} \bibinfo{volume}{24}
  (\bibinfo{year}{2017}) \bibinfo{pages}{58--66}.
\bibitem[{Jones et~al.(2007)Jones, Kumar, Pang, and Tomkins}]{jones:2007know}
\bibinfo{author}{R.~Jones}, \bibinfo{author}{R.~Kumar},
  \bibinfo{author}{B.~Pang}, \bibinfo{author}{A.~Tomkins},
\newblock \bibinfo{title}{I know what you did last summer: query logs and user
  privacy},
\newblock in: \bibinfo{booktitle}{Proceedings of the sixteenth ACM conference
  on Conference on information and knowledge management},
  \bibinfo{organization}{ACM}, pp. \bibinfo{pages}{909--914}.
\bibitem[{Li et~al.(2016)Li, Zhu, Du, Liang, and Shen}]{LiH:2016}
\bibinfo{author}{H.~Li}, \bibinfo{author}{H.~Zhu}, \bibinfo{author}{S.~Du},
  \bibinfo{author}{X.~Liang}, \bibinfo{author}{X.~Shen},
\newblock \bibinfo{title}{{Privacy Leakage of Location Sharing in Mobile Social
  Networks: Attacks and Defense}},
\newblock \bibinfo{journal}{IEEE Transactions on Dependable and Secure
  Computing}  (\bibinfo{year}{2016}) \bibinfo{pages}{1--1}.
\bibitem[{Joglekar and Kulkarni(2017)}]{Joglekar:2017wb}
\bibinfo{author}{P.~Joglekar}, \bibinfo{author}{V.~Kulkarni},
\newblock \bibinfo{title}{{Privacy Issues in Urban Computing using Mobile
  Crowdsensing}},
\newblock \bibinfo{journal}{International Journal of Computer Applications}
  \bibinfo{volume}{168} (\bibinfo{year}{2017}).
\bibitem[{To and Shahabi(2017)}]{To:2017vw}
\bibinfo{author}{H.~To}, \bibinfo{author}{C.~Shahabi},
\newblock \bibinfo{title}{{Location Privacy in Spatial Crowdsourcing}}
  (\bibinfo{year}{2017}).
\bibitem[{Enserink(2015)}]{Enserink:2015bz}
\bibinfo{author}{M.~Enserink},
\newblock \bibinfo{title}{{The End of Privacy}},
\newblock \bibinfo{journal}{Science} \bibinfo{volume}{347}
  (\bibinfo{year}{2015}) \bibinfo{pages}{490--491}.
\bibitem[{Ronen et~al.(2017)Ronen, Shamir, Weingarten, and
  O’Flynn}]{ronen:2017iot}
\bibinfo{author}{E.~Ronen}, \bibinfo{author}{A.~Shamir}, \bibinfo{author}{A.-O.
  Weingarten}, \bibinfo{author}{C.~O’Flynn},
\newblock \bibinfo{title}{{IoT goes nuclear: Creating a ZigBee chain
  reaction}},
\newblock in: \bibinfo{booktitle}{Security and Privacy (SP), 2017 IEEE
  Symposium on}, \bibinfo{organization}{IEEE}, pp. \bibinfo{pages}{195--212}.
\bibitem[{Pournajaf et~al.(2014)Pournajaf, Xiong, Garcia-Ulloa, and
  Sunderam}]{Pournajaf:2014wd}
\bibinfo{author}{L.~Pournajaf}, \bibinfo{author}{L.~Xiong},
  \bibinfo{author}{D.~A. Garcia-Ulloa}, \bibinfo{author}{V.~Sunderam},
  \bibinfo{title}{{A Survey on Privacy in Mobile Crowd Sensing Task
  Management}}, \bibinfo{type}{Technical Report} \bibinfo{number}{TR-2014-002},
  Mathematics and Computer Science, EMORY UNIVERSITY, \bibinfo{year}{2014}.
\bibitem[{Pournajaf et~al.(2016)Pournajaf, Garcia-Ulloa, Xiong, and
  Sunderam}]{Pournajaf:2016dr}
\bibinfo{author}{L.~Pournajaf}, \bibinfo{author}{D.~A. Garcia-Ulloa},
  \bibinfo{author}{L.~Xiong}, \bibinfo{author}{V.~Sunderam},
\newblock \bibinfo{title}{{Participant Privacy in Mobile Crowd Sensing Task
  Management: A Survey of Methods and Challenges}},
\newblock \bibinfo{journal}{ACM SIGMOD Record} \bibinfo{volume}{44}
  (\bibinfo{year}{2016}) \bibinfo{pages}{23--34}.
\bibitem[{Zhang et~al.(2017)Zhang, Ni, Yang, Liang, Ren, and
  Shen}]{ZhangK:2017}
\bibinfo{author}{K.~Zhang}, \bibinfo{author}{J.~Ni}, \bibinfo{author}{K.~Yang},
  \bibinfo{author}{X.~Liang}, \bibinfo{author}{J.~Ren}, \bibinfo{author}{X.~S.
  Shen},
\newblock \bibinfo{title}{{Security and Privacy in Smart City Applications:
  Challenges and Solutions}},
\newblock \bibinfo{journal}{IEEE Communications Magazine} \bibinfo{volume}{55}
  (\bibinfo{year}{2017}) \bibinfo{pages}{122--129}.
\bibitem[{Sinai et~al.(2014)Sinai, Partush, Yadid, and Yahav}]{aisrf:sinai2014}
\bibinfo{author}{M.~B. Sinai}, \bibinfo{author}{N.~Partush},
  \bibinfo{author}{S.~Yadid}, \bibinfo{author}{E.~Yahav},
\newblock \bibinfo{title}{Exploiting social navigation},
\newblock \bibinfo{journal}{arXiv preprint arXiv:1410.0151}
  (\bibinfo{year}{2014}).
\bibitem[{Blasco et~al.(2015)Blasco, Bustos-Jim{\'e}nez, Font, Hevia, and
  Prato}]{Blasco:2015ub}
\bibinfo{author}{S.~Blasco}, \bibinfo{author}{J.~Bustos-Jim{\'e}nez},
  \bibinfo{author}{G.~Font}, \bibinfo{author}{A.~Hevia}, \bibinfo{author}{M.~G.
  Prato},
\newblock \bibinfo{title}{{A three-layer approach for protecting smart-citizens
  privacy in crowdsensing projects.}},
\newblock \bibinfo{journal}{SCCC}  (\bibinfo{year}{2015}).
\bibitem[{Vergara-Laurens et~al.(2017)Vergara-Laurens, Jaimes, and
  Labrador}]{VergaraLaurens:2017hq}
\bibinfo{author}{I.~J. Vergara-Laurens}, \bibinfo{author}{L.~G. Jaimes},
  \bibinfo{author}{M.~A. Labrador},
\newblock \bibinfo{title}{{Privacy-Preserving Mechanisms for Crowdsensing:
  Survey and Research Challenges}},
\newblock \bibinfo{journal}{IEEE Internet of Things Journal}
  \bibinfo{volume}{4} (\bibinfo{year}{2017}) \bibinfo{pages}{855--869}.
\bibitem[{Huai et~al.(2015)Huai, Huang, Sun, and Yang}]{Huai:2015}
\bibinfo{author}{M.~Huai}, \bibinfo{author}{L.~Huang}, \bibinfo{author}{Y.-e.
  Sun}, \bibinfo{author}{W.~Yang},
\newblock \bibinfo{title}{{Efficient Privacy-Preserving Aggregation for Mobile
  Crowdsensing.}},
\newblock \bibinfo{journal}{BDCloud}  (\bibinfo{year}{2015}).
\bibitem[{Zhang et~al.(2016)Zhang, 0001, Gong, and Zhang}]{ZhangM:2016}
\bibinfo{author}{M.~Zhang}, \bibinfo{author}{L.~Y. 0001},
  \bibinfo{author}{X.~Gong}, \bibinfo{author}{J.~Zhang},
\newblock \bibinfo{title}{{Privacy-Preserving Crowdsensing - Privacy Valuation,
  Network Effect, and Profit Maximization.}},
\newblock \bibinfo{journal}{GLOBECOM}  (\bibinfo{year}{2016}).
\bibitem[{Vakilinia et~al.(2016)Vakilinia, Xin, Li, and Guo}]{Vakilinia:2016ur}
\bibinfo{author}{I.~Vakilinia}, \bibinfo{author}{J.~Xin},
  \bibinfo{author}{M.~Li}, \bibinfo{author}{L.~Guo},
\newblock \bibinfo{title}{{Privacy-Preserving Data Aggregation over Incomplete
  Data for Crowdsensing.}},
\newblock \bibinfo{journal}{GLOBECOM}  (\bibinfo{year}{2016}).
\bibitem[{Bakken et~al.(2004)Bakken, Rarameswaran, Blough, Franz, and
  Palmer}]{bakken:2004data}
\bibinfo{author}{D.~E. Bakken}, \bibinfo{author}{R.~Rarameswaran},
  \bibinfo{author}{D.~M. Blough}, \bibinfo{author}{A.~A. Franz},
  \bibinfo{author}{T.~J. Palmer},
\newblock \bibinfo{title}{Data obfuscation: Anonymity and desensitization of
  usable data sets},
\newblock \bibinfo{journal}{IEEE Security \& Privacy} \bibinfo{volume}{2}
  (\bibinfo{year}{2004}) \bibinfo{pages}{34--41}.
\bibitem[{Huning et~al.(2017)Huning, Bauer, and Aschenbruck}]{Huning:2017ve}
\bibinfo{author}{L.~Huning}, \bibinfo{author}{J.~Bauer},
  \bibinfo{author}{N.~Aschenbruck},
\newblock \bibinfo{title}{{A Privacy Preserving Mobile Crowdsensing
  Architecture for a Smart Farming Application.}},
\newblock \bibinfo{journal}{CrowdSenSys@SenSys}  (\bibinfo{year}{2017}).
\bibitem[{Dwork(2008)}]{dwork:2008differential}
\bibinfo{author}{C.~Dwork},
\newblock \bibinfo{title}{Differential privacy: A survey of results},
\newblock in: \bibinfo{booktitle}{International Conference on Theory and
  Applications of Models of Computation}, \bibinfo{organization}{Springer}, pp.
  \bibinfo{pages}{1--19}.
\bibitem[{Wang et~al.(2016)Wang, Zhang, Yang, Lim, and Ma}]{Wang:2016tu}
\bibinfo{author}{L.~Wang}, \bibinfo{author}{D.~Zhang},
  \bibinfo{author}{D.~Yang}, \bibinfo{author}{B.~Y. Lim},
  \bibinfo{author}{X.~Ma},
\newblock \bibinfo{title}{{Differential Location Privacy for Sparse Mobile
  Crowdsensing}},
\newblock \bibinfo{journal}{ICDM}  (\bibinfo{year}{2016}).
\bibitem[{Sei and Ohsuga(2016)}]{Sei:bu}
\bibinfo{author}{Y.~Sei}, \bibinfo{author}{A.~Ohsuga},
\newblock \bibinfo{title}{{Differential Private Data Collection and Analysis
  Based on Randomized Multiple Dummies for Untrusted Mobile Crowdsensing}},
\newblock \bibinfo{journal}{IEEE Transactions on Information Forensics and
  Security} \bibinfo{volume}{12} (\bibinfo{year}{2016})
  \bibinfo{pages}{926--939}.
\bibitem[{UID(2011)}]{UIDAI2011}
\bibinfo{title}{{UIDAI counters criticism over ‘false positives’}},
\newblock \bibinfo{journal}{Biometric Technology Today} \bibinfo{volume}{2011}
  (\bibinfo{year}{2011}) \bibinfo{pages}{3}.
\bibitem[{Bell and La~Padula(1976)}]{aisrf:bell1976}
\bibinfo{author}{D.~E. Bell}, \bibinfo{author}{L.~J. La~Padula},
  \bibinfo{title}{Secure computer system: Unified exposition and multics
  interpretation}, \bibinfo{type}{Report}, MITRE CORP BEDFORD MA,
  \bibinfo{year}{1976}.
\bibitem[{Bishop(2005)}]{aisrf:bishop2005}
\bibinfo{author}{M.~A. Bishop}, \bibinfo{title}{Introduction to computer
  security}, \bibinfo{publisher}{Addison-Wesley Professional},
  \bibinfo{year}{2005}.
\bibitem[{Denning(1976)}]{aisrf:denning1976}
\bibinfo{author}{D.~E. Denning},
\newblock \bibinfo{title}{A lattice model of secure information flow},
\newblock \bibinfo{journal}{Communications of the ACM} \bibinfo{volume}{19}
  (\bibinfo{year}{1976}) \bibinfo{pages}{236--243}.
\bibitem[{Kumar and Shyamasundar(2014)}]{aisrf:kumar2014}
\bibinfo{author}{N.~V.~N. Kumar}, \bibinfo{author}{R.~K. Shyamasundar},
\newblock \bibinfo{title}{Realizing purpose-based privacy policies succinctly
  via information-flow labels},
\newblock in: \bibinfo{booktitle}{2014 IEEE Fourth International Conference on
  Big Data and Cloud Computing (BdCloud)}, pp. \bibinfo{pages}{753--760}.
\bibitem[{Miller et~al.(2006)Miller, Engemann, and Yager}]{aisrf:miller2006}
\bibinfo{author}{H.~E. Miller}, \bibinfo{author}{K.~J. Engemann},
  \bibinfo{author}{R.~R. Yager},
\newblock \bibinfo{title}{Disaster planning and management},
\newblock \bibinfo{journal}{Communications of the IIMA} \bibinfo{volume}{6}
  (\bibinfo{year}{2006}) \bibinfo{pages}{25 -- 36}.
\bibitem[{UNISDR(2017)}]{aisrf:unisdr2017}
\bibinfo{author}{UNISDR}, \bibinfo{title}{Terminology
  ``\url{http://www.unisdr.org/we/inform/terminology}''}, \bibinfo{year}{2017}.
\bibitem[{{Centre for Research on the Epidemiology of Disasters
  (CRED)}(2016)}]{aisrf:cred2016}
\bibinfo{author}{{Centre for Research on the Epidemiology of Disasters
  (CRED)}}, \bibinfo{title}{2015 disasters in numbers}, \bibinfo{year}{2016}.
\bibitem[{{Asian Disaster Preparedness Center}(2011)}]{aisrf:asian2011}
\bibinfo{author}{{Asian Disaster Preparedness Center}}, \bibinfo{title}{{Module
  9: ICT for Disaster Risk Management}}, \bibinfo{year}{2011}.
\bibitem[{Bhatnagar et~al.(2016)Bhatnagar, Kumar, Ghosh, and
  Shyamasundar}]{aisrf:bhatnagar2016}
\bibinfo{author}{A.~Bhatnagar}, \bibinfo{author}{A.~Kumar},
  \bibinfo{author}{R.~K. Ghosh}, \bibinfo{author}{R.~K. Shyamasundar},
\newblock \bibinfo{title}{A framework of community inspired distributed message
  dissemination and emergency alert response system over smart phones},
\newblock in: \bibinfo{booktitle}{2016 8th International Conference on
  Communication Systems and Networks (COMSNETS)}, pp. \bibinfo{pages}{1--8}.
\bibitem[{Puthal et~al.(2015{\natexlab{a}})Puthal, Nepal, Ranjan, and
  Chen}]{aisrf:puthal2015a}
\bibinfo{author}{D.~Puthal}, \bibinfo{author}{S.~Nepal},
  \bibinfo{author}{R.~Ranjan}, \bibinfo{author}{J.~Chen},
\newblock \bibinfo{title}{Dpbsv -- an efficient and secure scheme for big
  sensing data stream},
\newblock in: \bibinfo{booktitle}{Internet of Things. IoT Infrastructures.
  IoT360 2015. Lecture Notes of the Institute for Computer Sciences, Social
  Informatics and Telecommunications Engineering}, pp.
  \bibinfo{pages}{246--253}.
\bibitem[{Puthal et~al.(2015{\natexlab{b}})Puthal, Nepal, Ranjan, and
  Chen}]{aisrf:puthal2015}
\bibinfo{author}{D.~Puthal}, \bibinfo{author}{S.~Nepal},
  \bibinfo{author}{R.~Ranjan}, \bibinfo{author}{J.~Chen},
\newblock \bibinfo{title}{A dynamic key length based approach for real-time
  security verification of big sensing data stream},
\newblock in: \bibinfo{editor}{edit} (Ed.), \bibinfo{booktitle}{Web Information
  Systems Engineering – WISE 2015}, volume \bibinfo{volume}{9419} of
  \textit{\bibinfo{series}{Lecture Notes in Computer Science}},
  \bibinfo{publisher}{Springer}, \bibinfo{year}{2015}{\natexlab{b}}, pp.
  \bibinfo{pages}{93--108}.
\bibitem[{Kumar and Shyamasundar(2016)}]{aisrf:kumar2016}
\bibinfo{author}{N.~V.~N. Kumar}, \bibinfo{author}{R.~K. Shyamasundar},
\newblock \bibinfo{title}{An end-to-end privacy preserving design of a
  map-reduce framework},
\newblock in: \bibinfo{booktitle}{2016 IEEE 18th International Conference on
  High Performance Computing and Communications; IEEE 14th International
  Conference on Smart City; IEEE 2nd International Conference on Data Science
  and Systems (HPCC/SmartCity/DSS)}, pp. \bibinfo{pages}{1469--1476}.
\bibitem[{Puthal et~al.(2017{\natexlab{a}})Puthal, Wu, Nepal, Ranjan, and
  Chen}]{aisrf:puthal2017}
\bibinfo{author}{D.~Puthal}, \bibinfo{author}{X.~Wu},
  \bibinfo{author}{S.~Nepal}, \bibinfo{author}{R.~Ranjan},
  \bibinfo{author}{J.~Chen},
\newblock \bibinfo{title}{Seen: A selective encryption method to ensure
  confidentiality for big sensing data streams},
\newblock \bibinfo{journal}{IEEE Transactions on Big Data}
  (\bibinfo{year}{2017}{\natexlab{a}}) \bibinfo{pages}{1--1}.
\bibitem[{Puthal et~al.(2017{\natexlab{b}})Puthal, Nepal, Ranjan, and
  Chen}]{aisrf:puthal2017a}
\bibinfo{author}{D.~Puthal}, \bibinfo{author}{S.~Nepal},
  \bibinfo{author}{R.~Ranjan}, \bibinfo{author}{J.~Chen},
\newblock \bibinfo{title}{A synchronized shared key generation method for
  maintaining end-to-end security of big data streams},
\newblock in: \bibinfo{booktitle}{Web Information Systems Engineering – WISE
  2015}.
\bibitem[{Zelazny(2012)}]{zelazny2012}
\bibinfo{author}{F.~Zelazny},
\newblock \bibinfo{title}{{The evolution of India’s UID program}},
\newblock \bibinfo{journal}{Center for Global Development}
  \bibinfo{volume}{168} (\bibinfo{year}{2012}).
\bibitem[{{OECD}(2007)}]{oecd:authentication2007}
\bibinfo{author}{{OECD}}, \bibinfo{title}{OECD Recommendation on Electronic
  Authentication and OECD Guidance for Electronic Authentication},
  \bibinfo{type}{Technical Report}, OECD, \bibinfo{year}{2007}.
\bibitem[{Schneider(2009)}]{aisrf:schneider}
\bibinfo{author}{F.~Schneider}, \bibinfo{title}{Authentication for people
  ``\url{https://www.cs.cornell.edu/fbs/publications/chptr.AuthPeople.pdf}''},
  \bibinfo{year}{2009}.
\bibitem[{Chase and Chow(2009)}]{Chase:2009}
\bibinfo{author}{M.~Chase}, \bibinfo{author}{S.~S. Chow},
\newblock \bibinfo{title}{Improving privacy and security in multi-authority
  attribute-based encryption},
\newblock in: \bibinfo{booktitle}{Proceedings of the 16th ACM Conference on
  Computer and Communications Security}, CCS '09.
\bibitem[{Li et~al.(2013)Li, Yu, Zheng, Ren, and Lou}]{Li2013}
\bibinfo{author}{M.~Li}, \bibinfo{author}{S.~Yu}, \bibinfo{author}{Y.~Zheng},
  \bibinfo{author}{K.~Ren}, \bibinfo{author}{W.~Lou},
\newblock \bibinfo{title}{Scalable and secure sharing of personal health
  records in cloud computing using attribute-based encryption},
\newblock \bibinfo{journal}{IEEE Transactions on Parallel and Distributed
  Systems} \bibinfo{volume}{24} (\bibinfo{year}{2013})
  \bibinfo{pages}{131--143}.
\bibitem[{Kuhn(2000)}]{kuhn2000}
\bibinfo{author}{D.~R. Kuhn}, \bibinfo{title}{Implementation of role-based
  access control in multi-level secure systems}, \bibinfo{year}{2000}.
  \bibinfo{note}{US Patent 6,023,765}.
\bibitem[{Sandhu et~al.(2000)Sandhu, Ferraiolo, Kuhn et~al.}]{sandhu2000}
\bibinfo{author}{R.~Sandhu}, \bibinfo{author}{D.~Ferraiolo},
  \bibinfo{author}{R.~Kuhn}, et~al.,
\newblock \bibinfo{title}{The nist model for role-based access control: towards
  a unified standard},
\newblock in: \bibinfo{booktitle}{ACM workshop on Role-based access control},
  volume \bibinfo{volume}{2000}, pp. \bibinfo{pages}{1--11}.
\bibitem[{Bertino et~al.(2001)Bertino, Bonatti, and Ferrari}]{bertino2001}
\bibinfo{author}{E.~Bertino}, \bibinfo{author}{P.~A. Bonatti},
  \bibinfo{author}{E.~Ferrari},
\newblock \bibinfo{title}{Trbac: A temporal role-based access control model},
\newblock \bibinfo{journal}{ACM Transactions on Information and System Security
  (TISSEC)} \bibinfo{volume}{4} (\bibinfo{year}{2001})
  \bibinfo{pages}{191--233}.
\bibitem[{Nemec et~al.(2017)Nemec, Sys, Svenda, Klinec, and
  Matyas}]{aisrf:rsa2048}
\bibinfo{author}{M.~Nemec}, \bibinfo{author}{M.~Sys},
  \bibinfo{author}{P.~Svenda}, \bibinfo{author}{D.~Klinec},
  \bibinfo{author}{A.~Matyas},
\newblock \bibinfo{title}{{The Return of Coppersmith’s Attack: Practical
  Factorization of Widely Used RSA Moduli}},
\newblock in: \bibinfo{booktitle}{IEEE CCS 2017}, pp.
  \bibinfo{pages}{1631--1648}.
\bibitem[{Vombatkere(2018)}]{aisrf:vombatkere}
\bibinfo{author}{S.~Vombatkere}, \bibinfo{title}{Data protection with concrete
  walls and uncrackable encryption,
  ``\url{https://countercurrents.org/2018/03/24/data-protection-with-concrete-walls-and-uncrackable-encryption/}''},
  \bibinfo{year}{24 March 2018}.
\bibitem[{Aadhaar(2018)}]{aisrf:aadhaarprotocol}
\bibinfo{author}{Aadhaar}, \bibinfo{title}{Aadhaar authentication api
  specification - version 2.0 (revision
  1)``\url{https://uidai.gov.in/images/FrontPageUpdates/aadhaar_authentication_api_2_0.pdf}''},
  \bibinfo{year}{Feb 2018}.
\bibitem[{Patil and Shyamasundar(2017)}]{aisrf:vtprks}
\bibinfo{author}{V.~T. Patil}, \bibinfo{author}{R.~K. Shyamasundar},
\newblock \bibinfo{title}{{Undoing of Privacy Policies on Facebook}},
\newblock in: \bibinfo{editor}{G.~Livraga}, \bibinfo{editor}{S.~Zhu} (Eds.),
  \bibinfo{booktitle}{{31th IFIP Annual Conference on Data and Applications
  Security and Privacy (DBSEC)}}, volume \bibinfo{volume}{LNCS-10359} of
  \textit{\bibinfo{series}{Data and Applications Security and Privacy XXXI}},
  \bibinfo{publisher}{{Springer International Publishing}},
  \bibinfo{address}{Philadelphia, PA, United States}, \bibinfo{year}{2017}, pp.
  \bibinfo{pages}{239--255}. \bibinfo{note}{Part 2: Privacy}.
\bibitem[{Dua et~al.(2009)Dua, Bulusu, Feng, and Hu}]{Dua:2009}
\bibinfo{author}{A.~Dua}, \bibinfo{author}{N.~Bulusu}, \bibinfo{author}{W.-C.
  Feng}, \bibinfo{author}{W.~Hu},
\newblock \bibinfo{title}{Towards trustworthy participatory sensing},
\newblock in: \bibinfo{booktitle}{Proceedings of the 4th USENIX Conference on
  Hot Topics in Security}, HotSec'09, pp. \bibinfo{pages}{8--8}.
\bibitem[{Huang et~al.(2010)Huang, Kanhere, and Hu}]{huang2010}
\bibinfo{author}{K.~L. Huang}, \bibinfo{author}{S.~S.~K. Kanhere},
  \bibinfo{author}{W.~Hu},
\newblock \bibinfo{title}{Are you contributing trustworthy data?: the case for
  a reputation system in participatory sensing},
\newblock in: \bibinfo{booktitle}{Proceedings of the 13th ACM international
  conference on Modeling, analysis, and simulation of wireless and mobile
  systems}, \bibinfo{organization}{ACM}, pp. \bibinfo{pages}{14--22}.
\bibitem[{Huang et~al.(2012)Huang, Kan, and Hu}]{huang2012}
\bibinfo{author}{K.~L. Huang}, \bibinfo{author}{S.~S. Kan},
  \bibinfo{author}{W.~Hu},
\newblock \bibinfo{title}{A privacy-preserving reputation system for
  participatory sensing},
\newblock in: \bibinfo{booktitle}{Local Computer Networks (LCN), 2012 IEEE 37th
  Conference on}, \bibinfo{organization}{IEEE}, pp. \bibinfo{pages}{10--18}.
\bibitem[{Christin et~al.(2013)Christin, Ro{\ss}kopf, Hollick, A, and
  Kanhere}]{christin2013}
\bibinfo{author}{D.~Christin}, \bibinfo{author}{C.~Ro{\ss}kopf},
  \bibinfo{author}{M.~Hollick}, \bibinfo{author}{L.~M. A},
  \bibinfo{author}{S.~S. Kanhere},
\newblock \bibinfo{title}{Incognisense: An anonymity-preserving reputation
  framework for participatory sensing applications},
\newblock \bibinfo{journal}{Pervasive and mobile Computing} \bibinfo{volume}{9}
  (\bibinfo{year}{2013}) \bibinfo{pages}{353--371}.
\bibitem[{He et~al.(2015)He, Chan, and Guizani}]{he2015}
\bibinfo{author}{D.~He}, \bibinfo{author}{S.~Chan},
  \bibinfo{author}{M.~Guizani},
\newblock \bibinfo{title}{User privacy and data trustworthiness in mobile crowd
  sensing},
\newblock \bibinfo{journal}{IEEE Wireless Communications} \bibinfo{volume}{22}
  (\bibinfo{year}{2015}) \bibinfo{pages}{28--34}.

\end{thebibliography}

\newpage

\end{document}